\centerline{\bf Fractional Operators in the Matrix Variate Case}

\vskip.3cm\centerline{A.M. MATHAI}
\vskip.2cm\centerline{Centre for Mathematical Sciences,}
\vskip.1cm\centerline{Arunapuram P.O., Pala, Kerala-68674, India, and}
\vskip.1cm\centerline{Department of Mathematics and Statistics, McGill University,}
\vskip.1cm\centerline{Montreal, Quebec, Canada, H3A 2K6}
\vskip.2cm\centerline{and}
\vskip.2cm\centerline{H.J. HAUBOLD}
\vskip.1cm\centerline{Office for Outer Space Affairs, United Nations}
\vskip.1cm\centerline{P.O. Box 500, Vienna International Centre} 
\vskip.1cm\centerline{A - 1400 Vienna, Austria, and}
\vskip.2cm\centerline{Centre for Mathematical Sciences,}
\vskip.1cm\centerline{Arunapuram P.O., Pala, Kerala-68674, India}

\vskip.3cm\centerline{\bf [This paper is dedicated to Professor Dr. Francesco Mainardi at his 70th birthday]}
\vskip.5cm\noindent{\bf Abstract}
\vskip.3cm Fractional integral operators connected with real-valued scalar functions of matrix argument are applied in problems of mathematics, statistics and natural sciences. In this article we start considering the case of a Gauss hypergeometric function with the argument being a rectangular matrix. Subsequently some fractional integral operators are introduced which complement these results available on fractional operators in the matrix variate cases. Several properties and limiting forms are derived. Then the pathway idea is incorporated to move among several different functional forms. When these are used as models for problems in the natural sciences then these can cover the ideal situations, neighborhoods, in between stages and paths leading to optimal situations.

Mathematical Subject Classification 2010: 26A33, 33C60, 33E12, 33E20

Key Words and Phrases: fractional calculus, fractional operators, multivariate and matrix-variate functions, Gauss hypergeometric function 

\vskip.3cm\noindent{\bf 1.\hskip.3cm Introduction}
\vskip.3cm The importance of anomalous reaction/relaxation and transport/diffusion is well recognized in many disciplines including physics, chemistry, biology, and engineering. Despite this fact, anomalous relaxation and transport are not well understood, and there is the need to develop mathematical and statistical models with predictive power. Recent developments of fractional calculus in terms of integro-differential operators that provide a unifying framework to model key aspects of anomalous relaxation and transport like non-locality, non-Markovian (memory) effects, and non-Gaussian (Levy) processes became available (cf. Mathai et al. [11], Nair [4]). The extension of such developments to matrix-variate statistical densities in general (Mathai and Haubold [8], Mainardi [13]) and to the specifically interesting case of Mittag-Leffler functions and matrix-variate analogues (Mathai [7]) have been achieved more recently. Such results are applicable to the solution of linear coupled fractional differential equations (Lim et al. [1]) as well as to the handling of fractional Poisson probability distributions (Laskin [12]). Further use of fractional integral operators connected with real-valued scalar functions of matrix argument can be utilized for fractional matrix calculus (Phillips [3]), numerical solution of fractional diffusion-wave equations (Garg and Manohar [6]), stability analysis of fractional-order systems (Jiao and Chen [9]), and probably for the application of the matrix-variate Mellin transform in radar image processing ( Anfinsen and Eltoft [10]).   
\vskip.3cm Let $X=(x_{ij})$ be a $p \times r,~r\ge p$ matrix of real distinct scalar variables $x_{ij}$'s. Let $A$ be a $p\times p$ real positive definite constant matrix, that is, $A=A'>O$, prime denoting the transpose. Let $B$ be a constant positive definite $r\times r$ matrix, $B>O$. Let $A^{1\over2}$ and $B^{1\over2}$ denote the positive definite square roots of $A$ and $B$ respectively. Let $X$ be of full rank $p$. Then $Z_X=A^{1\over2}XBX'A^{1\over2}$ is symmetric positive definite matrix. In this paper we will consider only real matrices. The corresponding results in the complex domain can be done parallel to the real case. The following standard notations will be used. Real matrix variate gamma will be denoted by
$$\Gamma_p(\alpha)=\pi^{{p(p-1)}\over4}\Gamma(\alpha)\Gamma(\alpha-{1\over2})...\Gamma(\alpha-{{j-1}\over2}),\eqno(1.1)
$$for $\Re(\alpha)>{{p-1}\over2}$ where $\Re(\cdot)$ denotes the real part of $(\cdot)$. It can be shown that $\Gamma_p(\alpha)$ has the integral representation
$$\Gamma_p(\alpha)=\int_{S>O}|S|^{\alpha-{{p+1}\over2}}{\rm e}^{-{\rm tr}(S)}{\rm d}S,~\Re(\alpha)>{{p-1}\over2}\eqno(1.2)
$$where $|S|$ means the determinant of the $p\times p$ positive definite matrix $S$ and ${\rm tr}(S)$ denotes the trace of $S$. The wedge product of the ${{p(p+1)}\over2}$ differentials ${\rm d}x_{ij}$'s will be denoted by
$${\rm d}S=\prod_{i\ge j}\wedge{\rm d}s_{ij},~\wedge =\hbox{ wedge}\eqno(1.4)
$$The type-1 beta integral is given by
$$B_p(\alpha,\beta)={{\Gamma_p(\alpha)\Gamma_p(\beta)}\over{\Gamma_p(\alpha+\beta)}}=\int_{O<X<I}|S|^{\alpha-{{p+1}\over2}}|I-S|^{\beta-{{p+1}\over2}}{\rm d}S,\eqno(1.5)
$$for $\Re(\alpha)>{{p-1}\over2},~\Re(\beta)>{{p-1}\over2}$, where $S$ is $p\times p$ positive definite and $O<S<I$ means $S>O,I-S>O$ or all the eigenvalues of $S$ are in the open interval $(0,1)$. In general, $\int_X$ means the integral over $X$. The type-2 beta integral is given by
$$\eqalignno{B_p(\alpha,\beta)&=\int_{S>O}|S|^{\alpha-{{p+1}\over2}}|I+S|^{-(\alpha+\beta)}{\rm d}S\cr
&=\int_{U>O}|U|^{\beta-{{p+1}\over2}}|I+U|^{-(\alpha+\beta)}{\rm d}U,\Re(\alpha)>{{p-1}\over2},\Re(\beta)>{{p-1}\over2}&(1.6)\cr}
$$Let
$$Z_X=A^{1\over2}XBX'A^{1\over2}\hbox{  and  }Z_Y=A^{1\over2}YBY'A^{1\over2}
$$where $X=(x_{ij})$ and $Y=(y_{ij})$ are $p\times r,r\ge p$ matrices of real elements, and of full rank $p$. Consider the evaluation of the integral
$$I_1=\int_X|Z_X|^{a}|I-Z_X|^{c-a-{{p+1}\over2}}|I-Z_YZ_X|^{-b}{\rm d}X,\eqno(1.7)
$$for $O<Z_X<I,O<Z_Y<I$. This integral corresponds to the Euler integral for Gauss hypergeometric function. Let $$U=A^{1\over2}XB^{1\over2}\Rightarrow {\rm d}U=|A|^{{r}\over2}|B|^{{p}\over2}{\rm d}X\hbox{ (see Mathai, 1997, equation (1.2.20)) }\eqno(1.8)
$$Then, after integration over the Stiefel manifold,
$$Z_X=UU'=V\Rightarrow {\rm d}U={{\pi^{{rp}\over2}}\over{\Gamma_p({{r}\over2})}}|V|^{{{r}\over2}-{{p+1}\over2}}{\rm d}V\eqno(1.9)
$$(see Mathai, 1997, Theorem 2.16 and Remark 2.13). Now, the integral $I_1$ becomes,
$$\eqalignno{I_1&=|A|^{-{{r}\over2}}|B|^{-{{p}\over2}}{{\pi^{{rp}\over2}}\over{\Gamma_p({{r}\over2})}}\cr
&\times
\int_{O<V<I}|V|^{a+{{r}\over2}-{{p+1}\over2}}|I-V|^{c-a-{{p+1}\over2}}|I-Z_Y^{1\over2}VZ_Y^{1\over2}|^{-b}{\rm
d}V.&(1.10)\cr}
$$Let us expand the last factor in the integrand in terms of zonal polynomials. For a discussion of zonal polynomials see Mathai et al. (1995).
$$|I-Z_Y^{1\over2}VZ_Y^{1\over2}|^{-b}=\sum_{k=0}^{\infty}\sum_K{{(b)_K}\over{k!}}C_K(Z_YV)\eqno(1.11)
$$where $C_K(\cdot)$ is the zonal polynomial of order $k$, $K=(k_1,k_2,...,k_p), k_1+...+k_p=k$ and
$$\eqalignno{(b)_K&=\prod_{j=1}^p(b-{{j-1}\over2})_{k_j}={{\Gamma_p(b,K)}\over{\Gamma_p(b)}},\cr
\Gamma_p(b,K)&=\pi^{{p(p-1)}\over4}\prod_{j=1}^p\Gamma(b+k_j-{{j-1}\over2})=\Gamma_p(b)(b)_K&(1.12)\cr}
$$and $(a)_m$ is the Pochhammer symbol
$$(a)_m=a(a+1)...(a+m-1), (a)_0=1, a\ne 0.
$$We can evaluate the integral
$$\int_{O<V<I}|V|^{a+{{r}\over2}-{{p+1}\over2}}|I-V|^{c-a-{{p+1}\over2}}C_K(Z_YV){\rm d}V={{\Gamma_p(a+{{r}\over2},K)\Gamma_p(c-a)}\over{\Gamma_p(c+{{r}\over2},K)}}C_K(Z_Y)\eqno(1,13)
$$(see Mathai, 1997, (5.1.26)). Note that interchange of integrals and sums is valid here. Then
$$\eqalignno{I_1&=|A|^{-{{r}\over2}}|B|^{-{{p}\over2}}{{\pi^{{rp}\over2}}\over{\Gamma_p({{r}\over2})}}
{{\Gamma_p(a+{{r}\over2})\Gamma_p(c-a)}\over{\Gamma_p(c+{{r}\over2})}}\cr
&\times \sum_{k=0}^{\infty}\sum_{K}{{(b)_K(a+{{r}\over2})_K}\over{(c+{{r}\over2})_K}}{{C_K(Z_Y)}\over{k!}}.&(1.14)\cr}$$

\vskip.3cm\noindent{\bf 2.\hskip.3cm Hypergeometric Functions of Rectangular Matrix Argument}

\vskip.3cm In the notations of hypergeometric function of matrix argument the series part of (1.14) can be written as a ${_2F_1}$. That is,
$$\sum_{k=0}^{\infty}\sum_K{{(b)_K(a+{{r}\over2})_K}\over{(c+{{r}\over2})_K}}{{C_K(Z_Y)}\over{k!}}
={_2F_1}(a+{{r}\over2},b;c+{{r}\over2};Z_Y),~\Vert Z_Y\Vert <1\eqno(2.1)
$$where $\Vert (\cdot)\Vert $ denotes a norm of $(\cdot)$. Hence we have the following theorem:

\vskip.3cm\noindent{\bf Theorem 2.1.}\hskip.3cm{\it For $X,Y,A,B$ as defined in Section 1
$$\eqalignno{{_2F_1}(a+{{r}\over2},&b;c+{{r}\over2};Z_Y)={{|A|^{{r}\over2}|B|^{{p}\over2}\Gamma_p({{r}\over2})
\Gamma_p(c+{{r}\over2})}\over{\pi^{{rp}\over2}\Gamma_p(a+{{r}\over2})\Gamma_p(c-a)}}\cr
&\times \int_{X}|Z_X|^{a}|I-Z_X|^{c-a-{{p+1}\over2}}|I-Z_YZ_X|^{-b}{\rm d}X&(2.2)\cr}
$$for $O<Z_X<I,O<Z_Y<I,\Re(c-a)>{{p-1}\over2},\Re(a)>-{{r}\over2}+{{p-1}\over2}$.}

 \vskip.2cm For establishing some limiting forms and pathways we need the following results, which will be stated as lemmas.

\vskip.3cm\noindent{\bf Lemma 2.1.}\hskip.3cm{\it For $K=(k_1,...,k_p), k_1+...+k_p=k, (a)_K=\prod_{j=1}^p(a-{{j-1}\over2})_{k_j}$,
$$\lim_{q\to 1}(q-1)^k({{1}\over{q-1}})_K=1.
$$}
\vskip.3cm Note that
$$\eqalignno{(q-1)^k({{1}\over{q-1}})_K&=\{\prod_{j=1}^p(q-1)^{k_j}\}\{\prod_{i=1}^{k_j}
({{1}\over{q-1}}-({{j-1}\over2})+i-1)\}\cr
&=\prod_{j=1}^p[1-(q-1)({{j-1}\over2})+(q-1)(i-1)]=1.\cr}$$

\vskip.3cm\noindent{\bf Lemma 2.2.}\hskip.3cm{\it
$$\lim_{q\to 1}|I+(q-1)Z_Y|^{-{{1}\over{q-1}}}={\rm e}^{-{\rm tr}(Z_Y)}.\eqno(2.4)$$}
\vskip.3cm\noindent{\bf Proof:}\hskip.3cm Let $\lambda_j,~j=1,...,p$ be the eigenvalues of $Z_Y$. Then
$$\eqalignno{|I+(q-1)Z_Y|&=\prod_{j=1}^p(1+(q-1)\lambda_j).\cr
\noalign{\hbox{But}} \prod_{j=1}^p[\lim_{q\to
1}\{1+(q-1)\lambda_j\}^{-{{1}\over{q-1}}}]&=\prod_{j=1}^p{\rm
e}^{-\lambda_j}={\rm e}^{-\sum_{j=1}^p\lambda_j}\cr &={\rm e}^{-{\rm
tr}(Z_Y)}.&(2.5)\cr}$$

\vskip.3cm\noindent{\bf 3.\hskip.3cm Fractional Integral Operators}

\vskip.3cm Let $Z_X$ and $Z_Y$ be as defined in Section 1. Let the generalized fractional integral operator of matrix argument be defined and denoted as

$$\eqalignno{({_0D}_X^{-\alpha}f)(X)&={{1}\over{\Gamma_p(\alpha)}}\int_{Z_X>Z_Y>O}|Z_X-Z_Y|^{\alpha-{{p+1}\over2}}f(Z_Y){\rm d}Y\cr
&={{|Z_X|^{\alpha-{{p+1}\over2}}}\over{\Gamma_p(\alpha)}}\int_{Z_X>Z_Y>O}
|I-Z_X^{-{1\over2}}Z_YZ_X^{-{1\over2}}|^{\alpha-{{p+1}\over2}}f(Z_Y){\rm d}Y.\cr}
$$Make the transformations $U=A^{1\over2}YB^{1\over2},~V=UU',~W=Z_X^{-{1\over2}}VZ_X^{-{1\over2}}$. Then integrating out over the Stiefel manifold we have
$$\eqalignno{{_0D}_X^{-\alpha}f&={{|Z_X|^{\alpha+{{r}\over2}-{{p+1}\over2}}}
\over{|A|^{{r}\over2}|B|^{{p}\over2}\Gamma_p(\alpha)}}{{\pi^{{rp}\over2}}\over{\Gamma_p({{r}\over2})}}\cr
&\times \int_{O<W<I}|I-W|^{\alpha-{{p+1}\over2}}f(Z_X^{1\over2}WZ_X^{1\over2})|W|^{{{r}\over2}-{{p+1}\over2}}{\rm d}W.&(3.1)\cr}
$$Consider the special cases of $f(\cdot)$. Consider the operator operating on a power function.

\vskip.3cm\noindent{\bf Case 1.}\hskip.3cm Let $f(Z_X)=|Z_X|^{\eta}$. Then
$$\eqalignno{{_0D}_X^{-\alpha}|Z_X|^{\eta}&={{|Z_X|^{\alpha+{{r}\over2}+\eta-{{p+1}\over2}}\pi^{{rp}\over2}}
\over{|A|^{{r}\over2}|B|^{{p}\over2}\Gamma_p(\alpha)\Gamma_p({{r}\over2})}}\int_{O<W<I}|W|^{{{r}\over2}+\eta-{{p+1}\over2}}|I-W|^{\alpha-{{p+1}\over2}}{\rm d}W\cr
&={{|Z_X|^{\alpha+{{r}\over2}+\eta-{{p+1}\over2}}\pi^{{rp}\over2}}\over{|A|^{{r}\over2}|B|^{{p}\over2}\Gamma_p({{r}\over2}})}{{\Gamma_p({{r}\over2}
+\eta)}\over{\Gamma_p(\alpha+{{r}\over2}+\eta)}},~\Re({{r}\over2}+\eta)>{{p-1}\over2}.&(3.2)\cr}
$$
\vskip.2cm\noindent Case 2.\hskip.3cm Zonal polynomial of order $k$, $f(Z_X)=C_K(Z_X)$
\vskip.3cm Then going through the same steps as above
$$\eqalignno{{_0D}_X^{-\alpha}C_K(Z_X)&={{|Z_X|^{\alpha+{{r}\over2}-{{p+1}\over2}}\pi^{{rp}\over2}}
\over{|A|^{{r}\over2}|B|^{{p}\over2}\Gamma_p(\alpha)\Gamma_p({{r}\over2})}}\int_{O<W<I}|W|^{{{r}\over2}-{{p+1}\over2}}\cr
&\times |I-W|^{\alpha-{{p+1}\over2}}C_K(Z_X^{1\over2}WZ_X^{1\over2}){\rm d}W,\hbox{   (Mathai, 1997, equation (5.1.26))}\cr
&={{|Z_X|^{\alpha+{{r}\over2}-{{p+1}\over2}}\pi^{{rp}\over2}}\over{|A|^{{r}\over2}|B|^{{p}\over2}\Gamma_p(\alpha)\Gamma_p({{r}\over2})}}{{\Gamma_p({{r}\over2},K)\Gamma_p(\alpha)}\over{\Gamma_p(\alpha+{{r}\over2},K)}}C_K(Z_X)\cr
&={{|Z_X|^{\alpha+{{r}\over2}-{{p+1}\over2}}\pi^{{rp}\over2}}\over{|A|^{{r}\over2}|B|^{{p}\over2}\Gamma_p(\alpha+{{r}\over2})}}{{({{r}\over2})_KC_K(Z_X)}\over{(\alpha+{{r}\over2})_K.}}\cr}
$$

\vskip.3cm\noindent{\bf 4.\hskip.3cm Extended Saigo Operators}

\vskip.3cm Let
$$f(Z_X)={_2F_1}(a,b;c;I-Z_X^{-{1\over2}}Z_YZ_X^{-{1\over2}})\phi(Z_Y),Z_X>Z_Y.
$$Here ${_2F_1}$ is a Gauss hypergeometric function of matrix argument $I-Z_X^{-{1\over2}}Z_YZ_X^{-{1\over2}}$ and $\phi(Z_Y)$ is an arbitrary function so that
$$\int_{Z_X>Z_Y}|I-Z_X^{-{1\over2}}Z_YZ_X^{-{1\over2}}|^{\alpha-{{p+1}\over2}}C_K(I-Z_X^{-{1\over2}}Z_YZ_X^{-{1\over2}})
\phi(Z_Y){\rm d}Y<\infty.
$$Opening up the ${_2F_1}$ in terms of zonal polynomials and then substituting $U=A^{1\over2}YBA^{1\over2},~ V=UU'$ we have

$$\eqalignno{&{_0D}_X^{-\alpha}[{_2F_1}(a,b;c;I-Z_X^{-{1\over2}}Z_YZ_X^{-{1\over2}})\phi(Z_Y)]\cr
&=\sum_{k=0}^{\infty}\sum_K{{(a)_K(b)_K}\over{k!(c)_K}}
{{|Z_X|^{\alpha-{{p+1}\over2}}\pi^{{rp}\over2}}\over{|A|^{{r}\over2}|B|^{{p}\over2}\Gamma_p(\alpha)
\Gamma_p({{r}\over2})}}\cr
&\int_{O<V<Z_X}|I-Z_X^{-{1\over2}}VZ_X^{-{1\over2}}|^{\alpha-{{p+1}\over2}}C_K(I-Z_X^{-{1\over2}}VZ_X^{-{1\over2}})
\phi(V)|V|^{{{r}\over2}-{{p+1}\over2}}{\rm d}V.&(4.1)\cr}
$$Consider the special case $\phi(V)=|V|^{\eta}$ and then make the transformation $Z_X^{-{1\over2}}VZ_X^{-{1\over2}}=W$. Then the right side of (4.1) reduces to the following:
$$\eqalignno{&\sum_{k=0}^{\infty}\sum_K{{(a)_K(b)_K}\over{k!(c)_K}}{{|Z_X|^{\alpha+{{r}\over2}+\eta-{{p+1}\over2}}
\pi^{{rp}\over2}}\over{|A|^{{r}\over2}|B|^{{p}\over2}\Gamma_p(\alpha)\Gamma_p({{r}\over2})}}
\int_{O<W<I}|I-W|^{\alpha-{{p+1}\over2}}|W|^{\eta+{{r}\over2}-{{p+1}\over2}}C_K(I-W){\rm d}W.\cr}$$The integral part reduces to the following:

$$\eqalignno{&\int_{O<W<I}|W|^{\eta+{{r}\over2}-{{p+1}\over2}}|I-W|^{\alpha-{{p+1}\over2}}C_K(I-W){\rm d}W\cr
&=\int_{O<T<I}|T|^{\alpha-{{p+1}\over2}}|I-T|^{\eta+{{r}\over2}-{{p+1}\over2}}C_K(T){\rm d}T\cr
&={{\Gamma_p(\alpha,K)\Gamma_p(\eta+{{r}\over2})}\over{\Gamma_p(\alpha+\eta+{{r}\over2},K)}}C_K(I)\cr
&={{\Gamma_p(\alpha)\Gamma_p(\eta+{{r}\over2})}\over{\Gamma_p(\alpha+\eta+{{r}\over2})}}
{{(\alpha)_K}\over{(\alpha+\eta+{{r}\over2})_K}}C_K(I).\cr}
$$Substituting back and denoting the left side by $I_X$ we have
$$\eqalignno{I_X&={{\pi^{{rp}\over2}}\over{|A|^{{r}\over2}|B|^{{p}\over2}\Gamma_p({{r}\over2})}}
{{\Gamma_p(\eta+{{r}\over2})}\over{\Gamma_p(\alpha+\eta+{{r}\over2})}}|Z_X|^{\alpha+\eta+{{r}\over2}-{{p+1}\over2}}\cr
&\times {_3F_2}(a,b,\alpha;c,\alpha+\eta+{{r}\over2};I).\cr}
$$

\vskip.3cm\noindent{\bf 5.\hskip.3cm  Some Statistical
Considerations} \vskip.3cm Let
$U_1=A_1^{1\over2}X_1B_1X_1'A_1^{1\over2},~U_2=A_2^{1\over2}X_2B_2X_2'A_2^{1\over2}$.
Let $X_1$ and $X_2$ be independently and exponentially distributed
random matrices where $A_1$ and $A_2$ are $p\times p$ real symmetric
positive definite constant matrices, $B_1$ is $r_1\times r_1$,
$r_1\ge p$ and $B_2$ is $r_2\times r_2,r_2\ge p$ constant positive
definite matrices, $X_1$ is $p\times r_1$ and $X_2$ is $p\times r_2$
matrices of distinct real scalar random variables and the matrices
be of full rank $p$, where a prime denotes the transpose and the
square roots are the unique positive definite square roots of $A_1$
and $A_2$ respectively. Let the densities of $X_1$ and $X_2$ be
denoted by $f_1(X_1)$ and $f_2(X_2)$ and the joint density, denoted
by $f(X_1,X_2)=f_1(X_1)f_2(X_2)$ due to independence. Further, let
$$f_j(X_j){\rm d}X_j=c_j{\rm e}^{-{\rm
tr}(A_j^{1\over2}X_jB_jX_j'A_j^{1\over2})}{\rm d}X_j,j=1,2\eqno(5.1)
$$where $c_j$ is the normalizing constant. Make the transformations
$V_j=A_j^{1\over2}X_jB_j^{1\over2}\Rightarrow {\rm
d}X_j=|A_j|^{-{{r_j}\over2}}|B_j|^{-{{p}\over2}}{\rm d}V_j$ where
${\rm d}X_j$ is the wedge product of all differentials in
$X_k=(x_{ij}^{(k)})$ or ${\rm
d}X_k=\prod_{i=1}^p\prod_{j=1}^{r_k}\wedge{\rm d}x_{ij}^{(k)}$,
$\wedge = $ wedge product. Let $W_j=V_jV_j'$ and integrate out over
the Stiefel manifold. Then we have
$${\rm
d}V_j={{\pi^{{r_jp}\over2}}\over{\Gamma_p({{r_j}\over2})}}|W_j|^{{{r_j}\over2}-{{p+1}\over2}}{\rm
d}W_j\eqno(5.2)
$$and ${\rm tr}(\cdot)$ denotes the trace of $(\cdot)$. Hence the
rectangular matrix $X_j$ having the exponential density means $W_j$
is having a real matrix-variate gamma density and further,
$${\rm
d}X_j={{\pi^{{r_jp}\over2}|W_j|^{{{r_j}\over2}-{{p+1}\over2}}}\over{|A_j|^{{r_j}\over2}|B_J|^{{p}\over2}\Gamma_p({{r_j}\over2})}}{\rm
d}W_j.\eqno(5.3)
$$If the densities of $V_j$ and $W_j$ are denoted by $g_j(V_j)$ and
$h_j(W_j)$ respectively then
$$\eqalignno{h_j(W_j){\rm d}W_j&={{1}\over{\Gamma_p({{r_j}\over2})}}|W_j|^{{{r_j}\over2}-{{p+1}\over2}}{\rm
e}^{-{\rm tr}(W_j)}{\rm d}W_j,W_j>O&(5.4)\cr g_j(V_j){\rm
d}V_j&={{1}\over{\pi^{{r_jp}\over2}}}{\rm e}^{-{\rm
tr}(V_jV_j')}{\rm d}V_j&(5.5)\cr f_j(X_j){\rm
d}X_j&={{|A_j|^{{r_j}\over2}|B_j|^{{p}\over2}}\over{\pi^{{r_jp}\over2}}}{\rm
e}^{-{\rm tr}(A_j^{1\over2}X_jB_jX_j'A_j^{1\over2})}{\rm
d}X_j.&(5.6)\cr}
$$

\vskip.3cm\noindent{\bf 5.1.\hskip.3cm Density of the sum}

\vskip.3cm Let us examine the density of the sum $U_1+U_1$. The
joint density of $X_1$ and $X_2$, denoted by $f(X_1,X_2)$, is given
by
$$\eqalignno{f(X_1,X_2){\rm d}X_1\wedge{\rm
d}X_2&=f_1(X_1)f_2(X_2){\rm d}X_1\wedge{\rm d}X_2.\cr
\noalign{\hbox{Make the transformations
$V_j=A_j^{1\over2}X_jB_j^{1\over2}, W_j=V_jV_j'$ then}}
f(X_1,X_2){\rm d}X_1\wedge{\rm
d}X_2&=f_1(V_1V_1')f_2(V_2V_2')\{\prod_{j=1}^2|A_j|^{-{{r_j}\over2}}|B_j|^{-{{p}\over2}}\}{\rm
d}V_1\wedge{\rm d}V_2\cr
&=f_1(W_1)f_2(W_2)\{\prod_{j=1}^2|A_j|^{-{{r_j}\over2}}|B_j|^{-{{p}\over2}}{{\pi^{{r_j}\over2}}\over{\Gamma_p({{r_j}\over2})}}|W_j|^{{{r_j}\over2}-{{p+1}\over2}}\}{\rm
d}W_1\wedge{\rm d}W_2.\cr}
$$Then $U_1+U_2=W_1+W_2$.
Make the transformation $U=W_1+W_2$ and $V=W_1$, the Jacobian is
unity and $W_1=V,W_2=U-V$ and the integration is over the positive
definite matrices $U$ and $V$ with $U-V>O$. If $X_1$ and $X_2$ are
independently distributed as in (5.1) then denoting the marginal
density of $U$ by $f^{*}(U)$ we have
$$\eqalignno{f^{*}(U){\rm
d}U&=\int_{U>V>O}f_1(U-V)f_2(V){\rm d}V&(5.7)\cr
&={{1}\over{\Gamma_p({{r_1}\over2})\Gamma_p({{r_2}\over2})}}|U-V|^{{{r_1}\over2}-{{p+1}\over2}}|V|^{{{r_2}\over2}-{{p+1}\over2}}{\rm
e}^{-{\rm tr}(U-V)-{\rm tr}(V)}{\rm d}V\wedge{\rm d}V.&(5.8)\cr
\noalign{\hbox{Make the transformation
$Z=U^{-{1\over2}}VU^{-{1\over2}}$ for fixed $U$, then}} f^{*}(U){\rm
d}U&={{|U|^{{{r_1+r_2}\over2}-{{p+1}\over2}}{\rm e}^{-{\rm
tr}(U)}}\over{\Gamma_p({{r_1}\over2})\Gamma_p({{r_2}\over2})}}\int_{Z>O}|Z|^{{{r_2}\over2}-{{p+1}\over2}}|I-Z|^{{r_1\over2}-{{p+1}\over2}}{\rm
d}Z\wedge{\rm d}U\cr &={{|U|^{{{r_1+r_2}\over2}-{{p+1}\over2}}{\rm
e}^{-{\rm tr}(U)}}\over{\Gamma_p({{r_1+r_2}\over2})}}{\rm d}U.\cr}
$$Thus the sum $U$ is  real matrix-variate gamma distributed. Note
that the integral in (5.7) is Riemann-Liouville left-sided
fractional integral for the power function or where the arbitrary
function is of the form $f(T)=|T|^{{{r_2}\over2}-{{p+1}\over2}}$.

\vskip.3cm\noindent{\bf Acknowledgment}

\vskip.3cm The authors would like to thank the Department of Science and Technology, Government of India, for the financial assistance for this work under Project No. SR/S4/MS:287/05.

\vskip.5cm\centerline{\bf References}

\vskip.5cm\noindent [1]\hskip.3cm Lim, S.C., Eab, C.H., Mak, K.H., Li, M., and Chen, S.Y. (2012): Solving linear coupled fractional differential equations by direct operational method and some applications {\it Mathematical Problems in Engineering}, doi:10.1155/2012/653939., 28 pages.

\vskip.3cm\noindent [2]\hskip.3cm Mathai, A.M. (1997): {\it Jacobians of Matrix Transformations and Functions of Matrix Argument}, World Scientific Publishing, New York.

\vskip.3cm\noindent [3]\hskip.3cm Phillips, P.C.B. (1987): Fractional matrix calculus and the distribution of multivariate tests {\it Time Series and Econometric Modeling}, Eds. I.B. MacNeill and G.J. Umphrey, D. Reidel Publishing Company, 219-234.

\vskip.3cm\noindent [4]\hskip.3cm Nair, S.S. (2009): Pathway fractional integration operator, {\it Fractional Calculus and Applied Analysis}, {\bf 12}, 237-252.

\vskip.3cm\noindent [5]\hskip.3cm Mathai, A.M., Provost, S.B., and Hayakawa, T. (1995): {\it Bilinear Forms and Zonal Polynomials}, Springer, New York.

\vskip.3cm\noindent [6]\hskip.3cm Garg, M. and Manohar, P. (2010): Numerical solution of fractional diffusion-wave equation with two space variables by matrix method, {\it Fractional Calculus and Applied Analysis}, {\bf 13}, 191-207.

\vskip.3cm\noindent [7]\hskip.3cm Mathai, A.M. (2010): Some properties of Mittag-Leffler functions and Matrix-variate analogues: A statistical perspective, {\it Fractional Calculus and Applied Analysis}, {\bf 13}, 113-132.

\vskip.3cm\noindent [8]\hskip.3cm Mathai, A.M. and Haubold, H.J. (2011): Matrix-variate statistical distributions and fractional calculus, {\it Fractional Calculus and Applied Analysis}, {\bf 14}, 138-155.

\vskip.3cm\noindent [9]\hskip.3cm Jiao, Z. and Chen Y.Q. (2011): Stability analysis of fractional-order systems with double non-commensurate orders for matrix case, {\it Fractional Calculus and Applied Analysis}, {\bf 14}, 436-453.

\vskip.3cm\noindent [10]\hskip.3cm Anfinsen, S.N. and Eltoft, T. (2011): Application of the matrix-variate Mellin transform to analysis of polarimetric radar images, {\it IEEE Transactions on Geoscience and Remote Sensing}, {\bf 49}, 2281-2295.

\vskip.3cm\noindent [11]\hskip.3cm Mathai, A.M., Saxena, R.K., and Haubold, H.J. (2010): {\it The H-Function: Theory and Applications}, Springer, New York.

\vskip.3cm\noindent [12]\hskip.3cm Laskin, N. (2009): Some applications of the fractional Poisson probability distribution, {\it Journal of Mathematical Physics}, {\bf 50}, 113513.

\vskip.3cm\noindent [13]\hskip.3cm Mainardi, F. (2010): {\it Fractional Calculus and Waves in Linear Viscoelasticity}, Imperial College Press, London.

\bye